\documentclass[pra,tighten,article,twocolumn,showpacs,amssymb,amsmath,latexsym]{revtex4}

\usepackage{graphicx}

\newcommand{\rE}{{\mbox{\rm E}}}

\begin{document}
\title{Prior entanglement between senders 
enables \\
perfect quantum network coding 
with modification}
\author{Masahito Hayashi}
\affiliation{ERATO-SORST Quantum Computation and Information Project,
Japan Science and Technology Agency,
201 Daini Hongo White Bldg.
5-28-3, Hongo, Bunkyo-ku, Tokyo 113-0033, Japan.
\email{masahito@qci.jst.go.jp}}


\begin{abstract}
We find a protocol 
transmitting two quantum states crossly in the butterfly network
only with prior entanglement between two senders.
This protocol requires only one qubit transmission or 
two classical bits transmission in each channel in the butterfly network.
It is also proved that
it is impossible without prior entanglement.
More precisely, an upper bound of average fidelity is given  
in the butterfly network when prior entanglement is not allowed.
\end{abstract}
\maketitle

\section{Introduction}
\label{sec:intro}
\begin{figure}[htbp]
\begin{center}
\includegraphics[width=8cm]{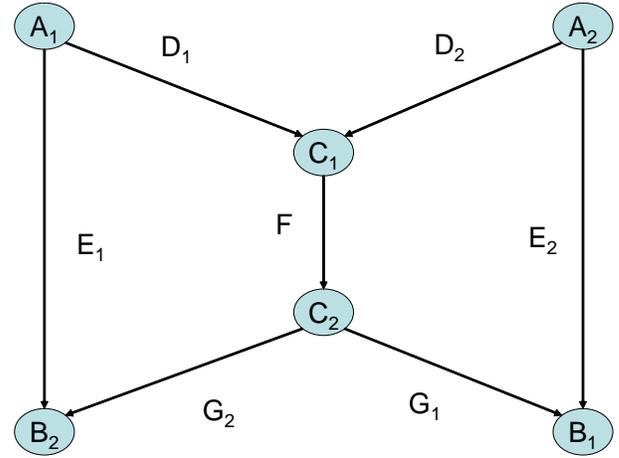}
\end{center}
\caption{Butterfly network}
\label{gb}
\end{figure}

Recently, long distance transmission of quantum state
has been actively researched by many various groups.
Hence, when global network of quantum communication becomes realized,
the efficient use of quantum network is essential. 
Especially, a large scale network often has a bottleneck point
that causes a transmission rate relatively small 
for the size of its communication resource.
Hence, it is required to resolve this bottleneck problem
in order to realize the high communication rate.
In the classical network system, 
Ahlswede et al.\cite{ACLY} 
formulated this problem as network coding,
and gave its solution as a coding in 
typical examples like the butterfly network (Fig \ref{gb}).
That is, they showed that the two informations can be sent crossly
in the butterfly network (Fig \ref{gb}), in which, 
all channels are allowed to transmit only one bit, then,
the bottleneck is the channel $F$.
The butterfly network seems only a specific example of network coding,
however, it represents the properties of networks with bottlenecks
so that their solution gave a trigger for a more general solution.
Concerning quantum system, Hayashi et al. \cite{HINRY}
initiated to study 
transmitting quantum state based on the quantum network
as its quantum extension.
In particular, they focused on the butterfly network,
and proved that perfect quantum state transmission is impossible 
in the butterfly network, i.e., the bottleneck $F$ cannot be resolved
in the quantum setting.
After this research, Iwama et al.\cite{INRY} treated 
quantum network coding with various types of networks.


On the other hand, prior entanglement provides some miracle performances 
in quantum information.
In dense coding\cite{dense}, 
prior entanglement enables the two-bit classical 
information only by one qubit transmission.
In quantum teleportation \cite{Tel}, prior entanglement enables 
the transmission of quantum state only by the sending classical information.
Hence, it is an interesting problem to discuss whether
prior entanglement enhances quantum network coding.

In order to resolve this bottleneck,
Leung et al.\cite{LOW} proposed a network code that 
transmits the quantum state 
crossly by use of shared entanglement between any two parties.
That is, they pointed out that
if all pairs of the sender and the receiver 
share prior entanglement,
combination of quantum teleportation and dense coding
enables perfect quantum state transmission 
in this network.
If two-bit transmission is allowed instead of 
one qubit transmission,
only using quantum teleportation,
perfect trasmission of quantum state is available.
However, 
their protocol requires preparing shared entanglement among four players.
So, it is required to reduce the number of players
sharing the prior entanglement
because increase of this number yields 
increase of the communication cost in the preparation stage.

In this paper, we treat the butterfly network,
with/without prior entanglement between only two senders.
As our result, we find that
prior entanglement between two senders
enables perfect quantum transmission
while it is impossible without prior entanglement 
even in the following modification of the rule of quantum network coding
treated by Hayashi et al. \cite{HINRY}
In this paper, 
we allow either one qubit transmission 
or two-bit classical communication in Fig \ref{gb}
although
Hayashi et al. \cite{HINRY} allow only one qubit transmission in all channels
in Fig \ref{gb}.
This is because one qubit transmission
can be exchanged with two-bit classical communication
under the prior entanglement.
We prove that 
perfect quantum state transmission is impossible even in such a modification.
This type of use of prior entanglement seems to suggest
further application of prior entanglement.
Hence, future research of this direction can be expected.

\begin{figure}[htbp]
\begin{center}
\includegraphics[width=8cm]{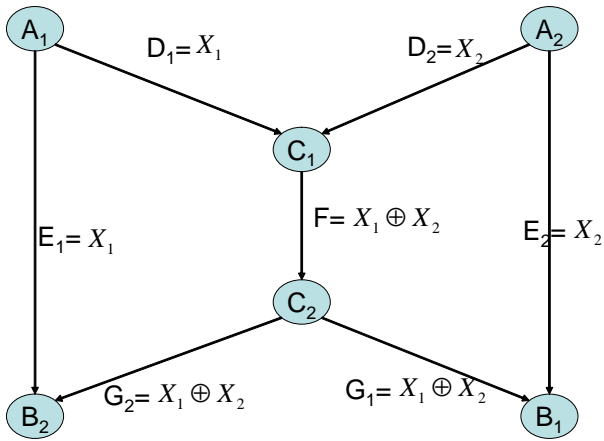}
\end{center}
\caption{Ahlswede et al. \cite{ACLY}'s protocol}
\label{gb2}
\end{figure}

Now, we give the classical protocol by Ahlswede et al. \cite{ACLY}
for the butterfly network (Fig \ref{gb}).
The purpose is 
sending the one-bit classical information $X_1$ 
from the site $A_1$ to the site $B_1$,
and
sending the other one-bit classical information $X_2$ 
from the site $A_2$ to the site $B_2$.
In this case,
all channels can send only one-bit information.
The bottleneck is the channel $F$ between the sites $C_1$ and $C_2$.
The solution is given by FIG \ref{gb2}.
That is, the receiver $B_i$ recovers the information $X_i$ by taking the sum 
of two received bits.

However, the same protocol is impossible in the quantum case.
In this paper, 
all channels 
$D_1$, $D_2$, $E_1$, $E_2$, $F$, $G_1$, $G_2$
can transmit only one qubit or two-bit classical information.
The purpose is 
sending the one-qubit state $\psi_1$ 
from the site $A_1$ to the site $B_1$,
and
sending the other one-qubit state $\psi_2$ 
from the site $A_2$ to the site $B_2$.
The main result is the protocol 
transmitting these two quantum states 
crossly with two prior Bell states between the senders $A_1$ and $A_2$.
Further, we prove its impossibility without prior entanglement.
That is, 
the average $\frac{f_1+f_2}{2}$ 
is less than 0.9504, 
where 
$f_i$ is the average fidelity between the sent state 
on $A_i$ and the recovered state on $B_i$
with the uniform distribution concerning $\psi_i$.
In our proof, we only use the constraint that 
the size of the channel $F$ is
either one qubit or two classical bits.
Other constraints are not essential for our proof of impossibility part.
Note that
Hayashi et al. \cite{HINRY} obtained the upper bound 
$0.983$ of fidelity. However, they concern the worst case instead of 
the average case.

Finally, we should comment the relation between the main idea and the
preceding researches.
Leung et al. \cite{LOW} gave the relationship 
between the secret sharing \cite{IMNTW}
and the capacity region of quantum network coding with the butterfly network.
Our proof of impossible part is motivated by this method.
However, our proof does not use any result concerning secret sharing.
Only the relation 
$I(R:A)+I(R:B)\le 2H(R)$ in \cite{IMNTW}
is used,
where $I(R:A)$ is the mutual information $H(R)+H(A)-H(RA)$
and $H(A)$ is the von Neumann entropy of the system $A$.
In fact, they conjectured that
prior entanglement between neighboring parties 
cannot enhance the ability of quantum network code in the butterfly network.
As is mentioned in discussions, we can show this conjecture as a byproduct.
This is a great advantage of our method.

This paper uses several relations in quantum information,
in which the corresponding relation in the book \cite{H} is referred.
Further, while their approach can treat only the asymptotic case where the fidelity goes to $1$, our approach can treat the finite fidelity.

\section{Our protocol}
\begin{figure}[htbp]
\begin{center}
\includegraphics[width=8cm]{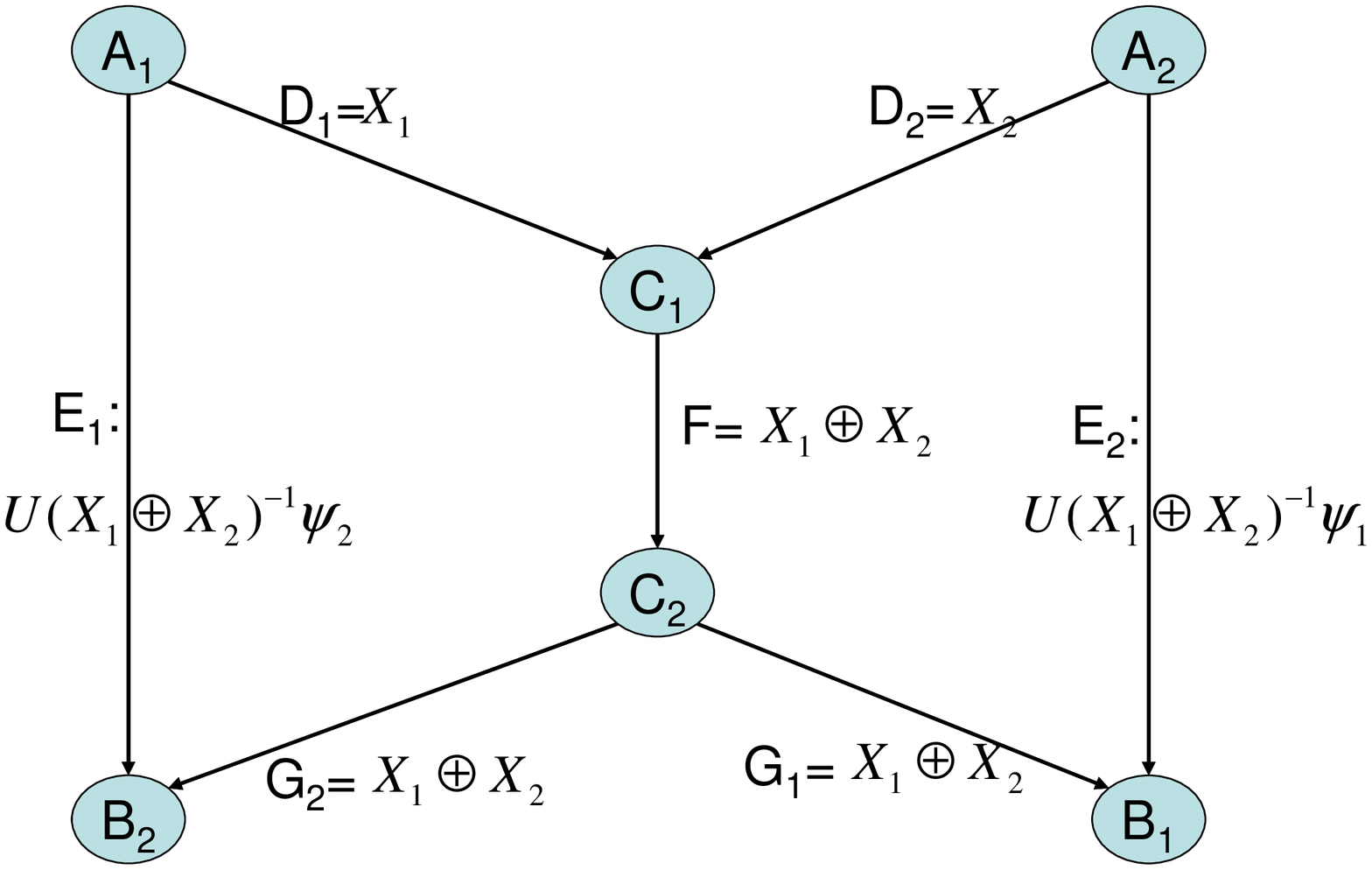}
\end{center}
\caption{Our protocol with prior entanglement}
\label{gb3}
\end{figure}

Now, we give our protocol, which enables transmitting 
quantum state perfectly and crossly based on the butterfly network,
whose protocol is summarized by Fig \ref{gb3}.
In our protocol, we essentially use quantum teleportation \cite{Tel}.
Assume that the two senders $A_1$ and $A_2$ share two pairs of 
the maximally entangled state $\Phi^+$,
where the first pair has two sites $A_{1,1}$ and $A_{2,1}$
and the second pair has other two sites $A_{1,2}$ and $A_{2,2}$.
The two senders $A_1$ and $A_2$ prepare 
their states $|\psi_1\rangle$ and $|\psi_2\rangle$.

In the first step, the sender $A_i$ performs the Bell measurement 
$\{\Phi^+,\Phi^-,\Psi^+,\Psi^-\}$
on the 
joint system $A_i \otimes A_{i,i}$, and obtains his data $X_i$,
where $(0,0)$, $(1,0)$, $(0,1)$, $(1,1)$ correspond to
$\Phi^+,\Phi^-,\Psi^+,\Psi^-$, respectively.
In this case the state on the remaining site $A_{i, i \oplus 1}$
is $U(X_{i\oplus 1})^{-1}|\psi_{i \oplus 1}\rangle$,
where $U(X)$ is the recovering unitary operation for teleportation 
with the outcome $X$.

In the second step, 
the sender $A_i$ performs the unitary operation $U(X_i)^{-1}$
to the remaining site $A_{i, i \oplus 1}$,
Hence, the state on the system $A_{i, i \oplus 1}$
becomes
$U(X_i)^{-1}U(X_{i\oplus 1})^{-1}|\psi_{i \oplus 1}\rangle=
c(X_i,X_{i\oplus 1})U(X_1\oplus X_2)^{-1}|\psi_{i \oplus 1}\rangle$,
where $|c(X_i,X_{i\oplus 1})|=1$.
Then, the sender $A_i$ sends the system $A_{i, i \oplus 1}$
to $B_{i \oplus 1}$ via the channel $E_i$.
He also sends the classical information $X_i$ via the channel $D_i$.
In the third step, 
the site $C_1$ sends the classical information $X_1\oplus X_2$
via the channel $F$.
Also, the site $C_2$ sends the same classical information
$X_1\oplus X_2$ to $B_1$ and $B_2$ via the channels $G_1$ and $G_2$.

In the final step, the receiver $B_i$ performs the unitary operation 
$U(X_1\oplus X_2)$ 
to the received state $U(X_1\oplus X_2)^{-1}|\psi_i\rangle$.
Then, he recovers the original state 
$U(X_1\oplus X_2)U(X_1\oplus X_2)^{-1}|\psi_i\rangle
=|\psi_i\rangle$.
This protocol can be extended to the qudit case.

One may think that the phase factor causes dephasing when the state to be sent is entangled.
Assume that $|\phi\rangle$ is the initial state among $A_1$, $A_2$, and their reference system $R$.
When the data $X_1$ and $X_2$ are obtained,
the final state is $
U(X_1\oplus X_2)U(X_1)^{-1}U(X_{2})^{-1}\otimes
U(X_1\oplus X_2)U(X_2)^{-1}U(X_{1})^{-1}\otimes
I_R
|\phi\rangle
=
c(X_1,X_{2})c(X_2,X_{1})|\phi\rangle$. Then, the state can be recovered perfectly.

\section{Properties of quantum information system}
For our proof of impossibility without prior entanglement,
we focus on the following six properties in quantum information system:
The number (*.*) represents the number of equation in \cite{H}.

{\it [P1] Monotonicity of quantum mutual information (8.38)
$I(A:B):= H(A)+H(B)-H(AB)$:}
\begin{align*}
I(R_1 R_2: B_1 B_2) \le I(R_1 R_2: E_1 E_2 F) .
\end{align*}

{\it [P2] Sum of quantum mutual information:}
As was shown by Imai et al.\cite{IMNTW},
the inequality
\begin{align*}
I(R_1:E_1)+I(R_1:B_1)\le 2 H(R_1)
\end{align*}
holds.

{\it [P3] Chain rule of quantum mutual information (5.75):}
The conditional quantum mutual information
$I(A:B|C):=H(AC)+H(BC)-H(ABC)-H(C)$ satisfies 
\begin{align*}
I(A:BC)= I(A:B|C)+I(A:C).
\end{align*}

{\it [P4] Convexity of quantum transmission information (8.42):}
For a channel $\kappa$ from $A$ to $B$,
the quantum transmission information $I_\kappa$ is defined by
$I(\kappa) := I(B:R)$ for the state 
$\kappa \otimes \iota_R (|\Phi\rangle \langle \Phi|)$,
where  
$|\Phi\rangle$ is the maximally entangled state between 
the input system and the reference system $R$.
Then, the convexity:
\begin{align*}
\lambda I(\kappa_1)+ (1-\lambda) I(\kappa_2) \ge 
I(\lambda \kappa_1+ (1-\lambda) \kappa_2) 
\end{align*}
holds.

{\it [P5] Quantum Fano inequality (8.51):}
The entanglement fidelity $f_{e}$
concerning the channel $\kappa$ from $A$ to $B$,
is given by 
$\langle \Phi^+ | 
\kappa\otimes \iota (|\Phi^+ \rangle \langle \Phi^+ | )
|\Phi^+ \rangle$.
This quantity satisfies 
\begin{align*}
H_{TW}(R B) \le \eta(f_{e}),
\end{align*}
where $R$ is the reference system, and 
$\eta(x):= -x \log_2 x - (1-x)\log_2\frac{1-x}{3}$.

{\it [P6] Twirling of channel:}
For any channel $\kappa$, 
its twirling $\overline{\kappa}$ is defined by
$\overline{\kappa}(\rho)
:= \int U \kappa (U \rho U^\dagger) U^\dagger p(d U)$,
where $p(d U)$ is the invariant distribution on $SU(2)$.
From the convexity,
the quantum transmission information 
of the original channel $\kappa$
is greater than that of its twirling $\overline{\kappa}$.
Further,
the entanglement fidelity of $\kappa$ is equal to 
that of $\overline{\kappa}$.

\section{Impossibility without prior entanglement}\label{s4}
In this section, we prove that the perfect quantum state transmission is 
impossible 
in the butterfly network without prior entanglement.
For this purpose, we will prove the entanglement fidelity $f_{e,i}$
concerning the channel $\kappa_i$ from $A_i$ to $B_i$
satisfies the following inequality:
\begin{align}
\frac{1}{2} \le \eta(\frac{f_{e,1}+f_{e,2}}{2}).
\label{4-16-1}
\end{align}
Solving this inequality yields
$\frac{f_{e,1}+f_{e,2}}{2} \le 0.9256$.
Since the average $f_i$ 
of the fidelity $\rE_{\psi_i}
\langle \psi_i| \kappa_i(|\psi_i\rangle \langle \psi_i|)|\psi_i\rangle 
$ is equal to $\frac{1+2 f_{e,i}}{3}$\cite{H},
we obtain $\frac{f_1+f_2}{2}\le 0.9504$.

Now, we prove the inequality (\ref{4-16-1}).
Let $R_i$ be the reference system of $A_i$.
We focus on the channel with the input $A_1$ and the output 
$E_1\otimes B_1$.
Since $H(R_1)=2$, P2 implies that
\begin{align}
I(R_1:E_1)\le 2- I(R_1:B_1).\label{4-16-2}
\end{align}
Similarly, 
\begin{align}
I(R_2:E_2)\le 2- I(R_2:B_2).\label{4-16-3}
\end{align}

As there is no prior entanglement,
the system $R_1 E_1 D_1 $ is independent of the system $R_2 E_2 D_2$.
Hence,
\begin{align}
&I(R_1 : R_2),I(E_1: E_2)
\le I(R_1 E_1: R_2 E_2) \nonumber \\
\le &I(R_1 E_1 D_1: R_2 E_2 D_1)=0.\label{4-17-1}
\end{align}
Thus, 
\begin{align}
&I(R_1 R_2: E_1 E_2) - I(R_1: E_1) - I(R_2:E_2) \nonumber \\
=&H(R_1 R_2)+H(E_1 E_2) -H(R_1 R_2 E_1 E_2) \nonumber \\
&- H(R_1)-H(E_1)+H(R_1 E_1)\nonumber \\
&- H(R_2)-H(E_2)+H(R_2 E_2)\nonumber \\
=&-I(R_1:R_2)-I(E_1:E_2)+
I(R_1 E_1: R_2 E_2)=0 \label{4-16-4}
\end{align}

When $F$ is a one-qubit quantum channel,
the relation $H(F E_1 E_2) \le H(E_1 E_2) + H(F)$ holds.
Now, let $H$ be the auxiliary system of $F R_1 R_2 E_1 E_2$,
then,
$H(R_1 R_2 E_1 E_2) -H(F R_1 R_2 E_1 E_2) 
=H(F H)- H(H)\le H(F)$.
Hence, 
\begin{align}
&I(R_1 R_2: F|E_1 E_2) \nonumber \\
=& 
H(R_1 R_2 E_1 E_2) 
-H(F R_1 R_2 E_1 E_2) \nonumber \\
&+H(F E_1 E_2) 
-H(E_1 E_2) \nonumber \\
\le & 2 H(F)\le 2.\label{4-16-5}
\end{align}

When $F$ is a two-bit classical channel,
the relation $H(R_1 R_2 E_1 E_2) \le H(F R_1 R_2 E_1 E_2) $
holds.
Thus,
\begin{align}
&I(R_1 R_2: F|E_1 E_2) \nonumber \\
= &
H(R_1 R_2 E_1 E_2) 
-H(F R_1 R_2 E_1 E_2) \nonumber \\
& +H(F E_1 E_2) 
-H(E_1 E_2) \nonumber \\
\le & H(F)\le 2.\label{4-16-6}
\end{align}

Combining the above relations with P3 and P1,
we obtain
\begin{align}
&I(R_1 R_2: B_1 B_2) \le
I(R_1 R_2: E_1 E_2 F) \nonumber \\
= &
I(R_1 R_2: F|E_1 E_2) 
+I(R_1 R_2: E_1 E_2) \nonumber \\
\le &
2+ I(R_1: E_1) + I(R_2:E_2) \nonumber \\
\le &
2+ 2- I(R_1:B_1)+2- I(R_2:B_2) .\label{4-16-7}
\end{align}
where the second inequality follows from
(\ref{4-16-4}) -- (\ref{4-16-6}),
and the final inequality follows from 
(\ref{4-16-2}) and (\ref{4-16-3}).

Now, we focus on the twirling of $\kappa_i$ by $\overline{\kappa}_i$,
and denote the transmission informations in the case of 
$\overline{\kappa}_1$, $\overline{\kappa}_2$,
and $\overline{\kappa}_1 \otimes \overline{\kappa}_2$
by 
$I_{TW}(R_1:B_1)$, $I_{TW}(R_2:B_2)$, and 
$I_{TW}(R_1 R_2: B_1 B_2)$, respectively.
In the case of the twirling channel,
the entropy of system $X$ is described by $H_{TW}(X)$.
Note that the entanglement fidelity of $\kappa_i$ is equal to 
that of $\overline{\kappa}_i$.
Using P6, we have
\begin{align}
& 6 \ge 
I(R_1 R_2: B_1 B_2) +
I(R_1:B_1)+I(R_2:B_2) \nonumber \\
\ge &
I_{TW}(R_1 R_2: B_1 B_2)+
I_{TW}(R_1:B_1)+I_{TW}(R_2:B_2) \nonumber \\
= &
4- H_{TW}(R_1 R_2 B_1 B_2) \nonumber \\
&+2- H_{TW}(R_1 B_1)
+2- H_{TW}(R_2 B_2) \nonumber \\
= &
8- 2 (H_{TW}(R_1 B_1)+H_{TW}(R_2 B_2)).\nonumber 
\end{align}
Thus, 
\begin{align}
1 \le H_{TW}(R_1 B_1)+H_{TW}(R_2 B_2).
\end{align}
P6 and P5 imply
\begin{align*}
H_{TW}(R_i B_i) \le \eta(f_{e,i}).
\end{align*}
Therefore, 
\begin{align*}
1 \le \eta(f_{e,1})+\eta(f_{e,2}).
\end{align*}
Using the convexity of $\eta$, we obtain (\ref{4-16-1})
.

In fact, our discussion on impossible part can be applied to 
the asymptotic case.
We assume that $N$ times use of the butterfly network,
$n_i$ qubit is sent from $A_i$ to $B_i$ perfectly.
As a generalization of (\ref{4-16-7}), the relation
\begin{align*}
&I(R_1 R_2: B_1 B_2) \\
\le &
2N + 2n_1- I(R_1:B_1)+2n_2- I(R_2:B_2)
\end{align*}
holds.
Since $I(R_i:B_i)\to 2n_i$,
and $I(R_1 R_2: B_1 B_2) \to 2n_1+2n_2$, we obtain
\begin{align*}
N \ge n_1+ n_2,
\end{align*}
which has been obtained by Leung et al.\cite{LOW}.

\section{Discussions}
In this paper, we focused on 
quantum network coding in the butterfly network,
and considered the effect of the existence of the prior entanglement 
between the senders $A_1$ and $A_2$.
As the first result, we found a protocol transmitting two quantum states 
crossly using two prior Bell states.
In the second result, we proved the impossibility when
no prior entangled state is allowed.
Our proof is based on information theoretical method,
while Hayashi et al. \cite{HINRY}'s evaluation is based on 
computational method.
In our proof, we use the non-existence of prior entanglement
only in (\ref{4-17-1}).
Other parts do not require this property.
Hence, even if the prior entanglement between neighboring parties in the network,
the discussion in Sec. \ref{s4} holds.
This argument was conjectured by Leung et. al. \cite{LOW}.

In our protocol with prior entanglement,
$I(R_1 E_1: R_2 E_2) =2$,
while $I(R_1: R_2) =I(E_1: E_2) =0$.
Hence, the property $I(R_1 E_1: R_2 E_2) =0$ 
is essential for our proof.
This fact indicates that good protocols should have non-zero 
mutual information $I(R_1 E_1: R_2 E_2)$.
That is, this fact may become a good 
indicator for seeking good network code in the quantum case.

\section*{Acknowledgments}
The author would like to thank Professor Hiroshi Imai of the
ERATO-SORST, QCI project for support.
He is grateful to Dr. Harumichi Nishimura and
Dr. Rudy Raymond for explaining the importance of 
the quantum network coding with prior entanglement.
They also informed him the essence of the presentation 
of Leung \cite{Leung}.
He also 
is grateful to Professor Keiji Matsumoto
for useful discussion.
He thanks Professor Samuel L. Braunstein and Dr. Peter Hines for comments 
concerning the phase factor.


\end{document}